\documentclass[final,5p,times,twocolumn,preprint]{elsarticle}
\usepackage{bm} 
\usepackage{graphicx}
\usepackage{epstopdf}
\usepackage{hyperref}
\usepackage{amsmath,amssymb,amsfonts,latexsym}
\allowdisplaybreaks
\begin{document}
\begin{frontmatter}
\title{Partonic angular momentum in the nucleon's chiral periphery} 
\tnotetext[t1]{JLAB Report JLAB-THY-19-2928}
\author[g]{C.~Granados}
\address[g]{Department of Physics, The George Washington University, Washington, DC 20052, USA}
\ead{cggranadosj@gmail.com}
\author[w]{C.~Weiss}
\address[w]{Theory Center, Jefferson Lab, Newport News, VA 23606, USA}
\ead{weiss@jlab.org}
\begin{abstract}
We study the nucleon's partonic angular momentum (AM) content at peripheral transverse distances 
$b = \mathcal{O}(M_\pi^{-1})$, where the structure is governed by chiral dynamics. We compute the nucleon 
form factors of the energy-momentum tensor in chiral effective field theory (ChEFT) and construct the 
transverse densities of AM at fixed light-front time. In the periphery the spin density is suppressed, 
and the AM is predominantly orbital. In the first-quantized representation of ChEFT in light-front form, 
the field-theoretical AM density coincides with the quantum-mechanical orbital AM density of the 
soft pions in the nucleon's periphery.
\end{abstract}
\end{frontmatter}
\section{Introduction}
Explaining how the nucleon spin emerges from the motion of quark and gluon fields is an essential 
step in understanding nucleon structure on the basis of QCD. The microscopic fields carry angular 
momentum (AM) both in their space-time dependence (orbital AM) and in their internal degrees 
of freedom (spin). The identification of these contributions is inherently not unique because
relativistic covariance and gauge symmetry connect both aspects of the motion.
Several decompositions of the QCD AM operator have been proposed, based on considerations of the 
mechanical interpretation of the operators, and of their relation to partonic operators that can be 
measured in deep-inelastic processes \cite{Jaffe:1989jz,Ji:1996nm}; 
see Ref.~\cite{Leader:2013jra} for a review. Thanks to extensive theoretical studies the correspondence
between these definitions is now well understood at the formal level; 
see Ref.~\cite{Lorce:2017wkb} for a recent summary. 
This concerns not just the global spin decomposition but also the local densities of 
spin and orbital AM, defined in terms of the axial vector current and the energy-momentum tensor
(EMT) operators \cite{Lorce:2017wkb,Polyakov:2002yz,Adhikari:2016dir}. 
It is therefore possible to use these concepts in dynamical calculations of nucleon structure.

The large-distance behavior of QCD is governed by the spontaneous breaking of chiral symmetry.
It leads to the appearance of Goldstone bosons, the pions, which are almost massless on the 
hadronic scale and mediate hadronic interactions at distances $\mathcal{O}(M_\pi^{-1})$.
The resulting dynamics can be formulated and solved using methods of chiral effective field 
theory (ChEFT) \cite{Gasser:1983yg,Weinberg:1991um}; see Refs.~\cite{Bernard:1995dp,Scherer:2012zzd} 
for a review. These methods can be used to compute nucleon structure at ``peripheral'' distances 
$\mathcal{O}(M_\pi^{-1})$ from first principles. A rigorous formulation of the spatial structure 
of the nucleon can be provided in the context of light-front (LF) quantization. 
The transverse densities at fixed LF time $x^+ \equiv x^0 + x^3$ are frame-independent 
(boost-invariant) and free of vacuum fluctuations, and permit an objective spatial representation of 
the nucleon as a relativistic system \cite{Burkardt:2002hr,Miller:2010nz}. ChEFT has been used
to calculate the transverse charge and current densities at peripheral distances
$b = \mathcal{O}(M_\pi^{-1})$, as well as the peripheral partonic structure; the results are 
model-independent and provide interesting insights into the mechanical properties of the nucleon at large 
distances \cite{Strikman:2010pu,Granados:2013moa,Granados:2015lxa,Granados:2015rra,Granados:2016jjl}.
It is thus natural to apply the same methods to the EMT and the AM densities in the nucleon.

In this letter we study the partonic AM in the nucleon's chiral periphery.
We calculate the nucleon form factors of the EMT in ChEFT and construct the transverse 
densities of AM at distances $b = \mathcal{O}(M_\pi^{-1})$. We find that at these distances
the spin density is suppressed and the AM density is predominantly orbital. Converting the
ChEFT expressions to a first-quantized representation in LF form, we show that the 
field-theoretical AM density coincides with the quantum-mechanical orbital AM density 
of the soft pions in the nucleon's periphery. As results of this study we (a) estimate 
the peripheral AM in the nucleon from first principles, (b) validate the concept of 
transverse AM density \cite{Lorce:2017wkb,Adhikari:2016dir}, (c)
obtain a simple mechanical picture of peripheral nucleon structure.
Details will be provided in a forthcoming article \cite{inprep}.

Our approach brings several new methodological elements to the study of partonic AM.
(a) ChEFT represents the ``true'' large-distance dynamics of QCD. The results therefore have universal
significance, in contrast to ad-hoc dynamical models. (b) ChEFT is a relativistic quantum field theory
and can illustrate aspects of field-theoretical AM besides its connection with QCD. 
ChEFT is essentially relativistic [the typical pion momenta are $k = \mathcal{O}(M_\pi)$] and requires
treatment of AM with methods appropriate for relativistic systems (LF formulation). 
ChEFT is not a gauge theory and allows us to explore field-theoretical AM without the
complications of gauge invariance. ChEFT is perturbative and has a limited particle content, making
it possible to change from a second-quantized, field-theoretical to a first-quantized, particle-based 
representation for the interpretation of the results.
\section{Transverse angular momentum densities}
In relativistic field theory the AM consists of an ``orbital'' and a ``spin'' part, resulting from 
rotations acting on the space-time dependence and the internal degrees of freedom of the fields.
In QCD the total AM can be grouped into gauge-invariant quark and gluon contributions. The gluon 
contribution cannot be split into orbital and spin parts in a gauge-invariant manner. Several 
definitions of quark and gluon AM have been proposed; see Refs.~\cite{Leader:2013jra,Lorce:2017wkb}
for a detailed discussion. In our study we use the ``kinetic'' definition \cite{Lorce:2017wkb}; 
our results for the peripheral densities do not depend on this choice, and the relation to other 
definitions is discussed below. The total (quark plus gluon) AM tensor of QCD is organized as
\begin{align}
J^{\mu\alpha\beta}(x) &= S_q^{\mu\alpha\beta}(x) + \widetilde{J}^{\mu\alpha\beta}(x) .
\label{spin_decomposition}
\end{align}
$S_q$ is the quark spin contribution and expressed in terms of the flavor-singlet 
axial current of the quark field,
\begin{align}
S_q^{\mu\alpha\beta}(x) &= \frac{1}{2} \epsilon^{\mu\alpha\beta\gamma} 
{\sum}_f \bar\psi_f(x) \gamma_\gamma \gamma^5 \psi_f(x) .
\label{S_q}
\end{align}
$\widetilde{J}$ is the ``rest'' of the AM and given in terms of the total (quark plus gluon) 
kinetic EMT of QCD,
\begin{align}
\widetilde{J}^{\mu\alpha\beta}(x) &= x^\alpha T^{\mu\beta}(x) - x^\beta T^{\mu\alpha}(x) .
\label{J_r_from_T}
\end{align}
$T$ can be split into gauge-invariant quark and gluon contributions \cite{Ji:1996nm}. Correspondingly, 
$\widetilde{J}$ can be split into a quark contribution describing the quark orbital AM, and a gluon 
contribution describing the gluon spin and orbital AM \cite{Lorce:2017wkb}. We do not require this 
splitting in the following and consider Eq.~(\ref{J_r_from_T}) with the total EMT of QCD (which can 
be matched with the EMT of ChEFT; see below). We refer to Eq.~(\ref{J_r_from_T}) in short as the 
``orbital'' AM, keeping in mind that it includes the gluonic contribution and is really the total 
AM ``other than quark spin.''

The kinetic EMT in Eq.~(\ref{J_r_from_T}) is not symmetric \cite{Lorce:2017wkb}. Its transition
matrix element between nucleon states with 4-momenta $p_{1, 2}$ and spin quantum numbers 
$\sigma_{1, 2}$ is parametrized as (at space-time point $x = 0$) \cite{Leader:2013jra,Bakker:2004ib}
\begin{align}
\left\langle N_2|T^{\mu\nu}|N_1\right\rangle &= 
\bar{u}_2\left[\gamma^{\{\mu}p^{\nu\}}A
- \frac{p^{\{\mu} \sigma^{\nu\}\alpha}\Delta_\alpha}{2M_N} \, B
\right.
\nonumber \\
&+ \frac{\Delta^\mu\Delta^\nu-\Delta^2 g^{\mu\nu}}{M_N} \, C
- \frac{p^{[\mu} \sigma^{\nu]\alpha}\Delta_\alpha}{2M_N} \, D
\nonumber \\
&+ \left. M_N g^{\mu\nu} \tilde{C} \right]u_1 ,
\label{emt_matrix_element}
\end{align}
where $p\equiv(p_1+p_2)/2$, $\Delta \equiv p_2-p_1$; $u_1 \equiv u(p_1, \sigma_1)$ etc.\ are the 
nucleon bispinors, $\sigma^{\mu\nu} \equiv \frac{1}{2} [\gamma^\mu, \gamma^\nu]$;
and we use $a^{\{\mu}b^{\nu\}}, a^{[\mu}b^{\nu]} \equiv \frac{1}{2}(a^\mu b^\nu \pm 
a^\nu b^\mu)$. The form factors $A$--$\tilde{C}$ are invariant functions of $t \equiv \Delta^2$. 
They satisfy the sum rules $A(0) = 1$ and $B(0) = 0$; other sum rules are discussed in 
Ref.~\cite{Polyakov:2018zvc}.

Local densities of AM have been defined based on the above operators and their form 
factors \cite{Lorce:2017wkb,Polyakov:2002yz,Adhikari:2016dir}.
Since the dynamical system is relativistic, it is natural to consider densities at fixed LF 
time $x^+ =$ const. In a frame where the momentum transfer is transverse 
$\Delta^\pm = 0, \, \bm{\Delta}_T \neq 0, \, t = -\bm{\Delta}_T^2$, one defines transverse densities 
associated with the invariant form factors $A$ and $B$ as
\begin{align}
\rho_{A}(b) &\equiv 
\int \frac{d^2\Delta_T}{(2\pi)^2} \, e^{-i\bm{\Delta}_T\cdot\bm{b}} \, A(t = -\bm{\Delta}^2_T),
\label{rho_A_rho_B}
\end{align}
and likewise for $B$. These densities are invariant under longitudinal boosts and 
free of vacuum fluctuation contributions to the matrix element \cite{Miller:2010nz}. 
In the context of the partonic description of nucleon structure, $\rho_A + \rho_B$ is equal to
the sum of the second moments ($x$-weighted integrals) of the generalized parton distributions
$H$ and $E$ in the impact parameter representation \citep{Ji:1996nm,Burkardt:2002hr}; 
as such this combination of densities is connected to measurable matrix 
elements of twist-2 QCD operators and of general interest. In a similar way one defines the
transverse density of the axial form factor associated with the nucleon matrix element of 
the axial current operator in Eq.~(\ref{S_q}) \cite{Lorce:2017wkb},
\begin{align}
\rho_{S}(b) &\equiv \int \frac{d^2\Delta_T}{(2\pi)^2} \, e^{-i\bm{\Delta}_T\cdot\bm{b}} 
\, G_A(t = -\bm{\Delta}^2_T).
\label{rho_S}
\end{align}
Transverse densities of AM at fixed LF time can be constructed from these invariant form 
factor densities \cite{Lorce:2017wkb,Adhikari:2016dir}.
To construct the ``orbital'' AM density associated with Eq.~(\ref{J_r_from_T}), 
we consider the EMT component $T^{+T}$ and take its matrix element between nucleon LF helicity 
states \cite{Brodsky:1997de} with LF helicities $\sigma_1 = \sigma_2 \equiv \sigma = \pm\frac{1}{2}$, 
corresponding to spin expectation value $S^z[\textrm{rest}] = \sigma$ in the nucleon rest frame,
\begin{align}
\langle \bm{T}^{+T}\rangle(\bm{\Delta}_T) &\equiv \langle p^+, \bm{\Delta}_T/2, \sigma | T^{+T} | 
p^+, -\bm{\Delta}_T/2, \sigma \rangle ,
\label{T_Delta}
\\[1ex]
\bm{T}^{+T}(\bm{b}) &= \int \frac{d^2\Delta_T}{(2\pi)^2} \, e^{-i\bm{\Delta}_T\cdot\bm{b}}
\; \langle \bm{T}^{+T} \rangle(\bm{\Delta}_T) .
\label{Tperpb}
\end{align}
The transverse density of orbital AM in the longitudinally polarized nucleon is then defined 
as \cite{Lorce:2017wkb}
\begin{align}
\langle L^z \rangle (b) &\equiv \frac{[\bm{b}\times \bm{T}^{+T}(\bm{b})]^z}{2p^+}
\label{L_z_from_T}
\\[.5ex]
 &= - \frac{\sigma}{2} \left( b \, \frac{d}{db} \right) (\rho_A + \rho_B + \rho_D)(b) .
\label{L_z_from_rho}
\end{align}
The expression Eq.~(\ref{L_z_from_rho}) is obtained by evaluating Eq.~(\ref{L_z_from_T}) with 
the specific form of the matrix element Eq.(\ref{emt_matrix_element}). $\rho_D$ is the transverse 
density associated with the form factor $D$, defined in analogy to Eq.~(\ref{rho_A_rho_B}).
The tensor in Eq.~(\ref{L_z_from_T}) and the densities in Eq.~(\ref{L_z_from_rho}) refer to the 
total EMT (sum of quark and gluon contributions). The transverse density of the quark spin associated 
with Eq.~(\ref{S_q}) is defined as
\begin{align}
\langle S^z \rangle (b) &\equiv \sigma \; \rho_S (b). 
\label{S_z}
\end{align}
Together the AM densities satisfy the sum rule \cite{Lorce:2017wkb}
\begin{align}
\int d^2b \, [ \langle S^z \rangle \, + \, \langle L^z \rangle ](b) \; = \; \sigma
\; = \; S^z[\textrm{rest}].
\end{align}

The densities defined by Eqs.~(\ref{L_z_from_T})--(\ref{S_z}) have a mechanical interpretation 
as AM densities in LF quantization and a simple relation to the invariant form factors of the EMT. 
This dual role allows them to be computed using various theoretical methods (including methods
requiring an invariant formulation, such as dispersion theory or Euclidean correlation functions) 
and makes them valuable tools for nucleon structure studies.
\section{Angular momentum in chiral periphery}
At peripheral transverse distances $b = \mathcal{O}(M_\pi^{-1})$ the EMT densities
Eq.~(\ref{rho_A_rho_B}) and the AM densities Eqs.~(\ref{L_z_from_T})--(\ref{S_z}) are governed by 
chiral dynamics and can be computed model-independently using ChEFT. The general method is based on 
the analytic properties of the form factors in $t$ and described in detail in 
Refs.~\cite{Granados:2013moa,Granados:2015rra,Granados:2016jjl}; here we discuss only the aspects 
specific to the AM operators.

When the effective large-distance dynamics of QCD is described by ChEFT, the total (quark plus gluon) 
EMT of QCD can be matched with the total EMT of ChEFT, 
\begin{align}
T^{\mu\nu}[\textrm{QCD}] \; \rightarrow \; T^{\mu\nu}[\textrm{ChEFT}],
\end{align}
as both operators arise from the invariance of the dynamics under space-time translations. 
The EMT of ChEFT with nucleons consists of terms involving the pion fields, the nucleon fields, 
and the pion-nucleon interactions \cite{Becher:1999he}. 
Contributions to the nucleon densities at peripheral distances 
$b = \mathcal{O}(M_\pi^{-1})$ are generated by chiral processes in which the operator couples 
to the nucleon by two-pion exchange (two-pion cut of the invariant form factors); processes in which
the operator couples directly to the nucleon result in contributions at short distances 
$b = \mathcal{O}(M_N^{-1})$. The relevant operator is therefore the pionic part of the EMT of ChEFT. 
At $\mathcal{O}(p^2)$ it is given by (the summation is over the isospin of the pion field)
\begin{align}
T^{\mu\nu}[\textrm{ChEFT, $\pi$}] \; =& \; {\sum}_a \left(
\partial^\mu \pi^a \partial^\nu \pi^a - {\textstyle\frac{1}{2}} g^{\mu\nu} 
\partial^\rho \pi^a \partial_\rho \pi^a \right.
\nonumber \\
&+ \left. {\textstyle\frac{1}{2}} g^{\mu\nu} M_\pi^2 \pi^a \pi^a \right) + \textrm{terms $\pi^4$, ...}
\label{EMT_chiral}
\end{align}
Its form is uniquely determined by chiral invariance and cannot be modified by adding terms that
are conserved without the equations of motion 
(improvement) \cite{Voloshin:1982eb,Leutwyler:1989tn,Polyakov:1999gs}.
In particular, the tensor Eq.~(\ref{EMT_chiral}) is symmetric. This implies that the peripheral
densities of the form factors $A$ and $B$ [related to the symmetric part of
Eq.~(\ref{emt_matrix_element})] are leading, while those of $D$ (related to the antisymmetric part)
are suppressed,
\begin{align}
\left.
\begin{array}{ll} \rho_{A, B}(b) & \textrm{leading} 
\\[1ex]
\rho_{D}(b) & \textrm{suppressed} 
\end{array}
\right\} &
\hspace{2em} 
[b = \mathcal{O}(M_\pi^{-1})] .
\end{align}
The same logic can be applied to the nucleon matrix element of the quark axial current in
Eq.~(\ref{S_q}). The isoscalar axial current in ChEFT does not have terms quadratic in the 
pion field; the form factor $G_A$ therefore does not have a two-pion cut; and the density
is suppressed at peripheral distances,
\begin{align}
\rho_S(b) \;\; \textrm{suppressed} 
& \hspace{2em} 
[b = \mathcal{O}(M_\pi^{-1})] .
\end{align}
Here we mean that the ``leading'' densities decay at large $b$ as $\sim\exp (-2 M_\pi b)$ 
(with a $b$-dependent pre-exponential factor), whereas the ``suppressed'' ones decay at 
least $\sim\exp (-3 M_\pi b)$. Altogether, we find that at peripheral distances the AM 
density is predominantly orbital,
\begin{align}
\langle L^z \rangle(b) \; \gg \; \langle S^z \rangle(b) 
& \hspace{2em} 
[b = \mathcal{O}(M_\pi^{-1})] ,
\end{align}
with the main contribution coming from $\rho_A$ and $\rho_B$ in Eq.~(\ref{L_z_from_rho}). 
This conclusion relies only on chiral symmetry, the quantum numbers of the AM operators, 
and the restriction to peripheral distances, and is therefore robust. 

It appears natural that the spin density $\rho_S$ is suppressed in the same way as the 
orbital AM density resulting from the antisymmetric part of the EMT, $\rho_D$. In fact,
the spin AM tensor $S_q$ in Eq.~(\ref{spin_decomposition}) is related to the antisymmetric 
part of the kinetic quark EMT by the condition of AM conservation \cite{Lorce:2017wkb}. 
Note that our conclusions would not change if we used the Belinfante instead of the 
kinetic definition of the EMT and the AM tensors \cite{Lorce:2017wkb}: the 
symmetric part of the EMT is the same in both definitions, and differences appear 
only in the antisymmetric parts of the EMT and the spin AM tensors, 
which are both suppressed at $b = \mathcal{O}(M_\pi^{-1})$.

\begin{figure}
\begin{center}
\includegraphics[width=0.84\columnwidth]{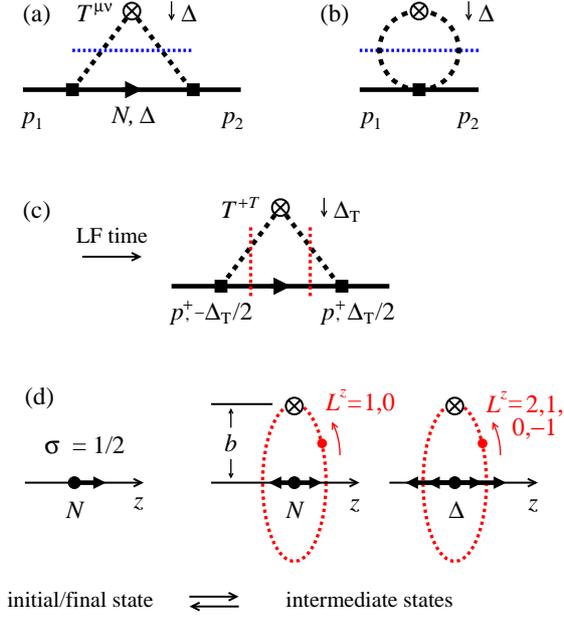}
\end{center}
\caption{(a)~LO ChEFT process contributing to the two-pion cut of the EMT form factors 
$A$ and $B$ (cut Feynman diagram). (b)~Contact term diagrams do not contribute to $A$ or $B$.
(c)~LF formulation of the ChEFT process (LF time-ordered diagram). (d)~Quantum-mechanical picture
of the ChEFT processes. The initial nucleon with spin $\sigma = 1/2$ makes a transition to a 
$\pi B$ intermediate state ($B = N, \Delta$). The peripheral AM density is generated by the 
orbital AM of the soft pion, $L^z$.} 
\label{fig:diag_EMT}
\end{figure}

We have calculated the leading peripheral EMT densities $\rho_{A, B}$ and the AM density 
$\langle L^z \rangle$ in ChEFT at LO. The calculation has been performed in two ways:
(I)~By computing the two-pion spectral functions of the invariant form factors $A$ and $B$ 
in the invariant formulation of ChEFT (cut Feynman diagrams) \cite{Becher:1999he}, and constructing 
the peripheral densities through a dispersion relation \cite{Granados:2013moa}. At LO the only 
diagram contributing to the two-pion cut of $A$ and $B$ is Fig.~\ref{fig:diag_EMT}a, 
where the vertices are the LO $\pi NN$ coupling. Diagrams with $\pi\pi NN$ contact terms 
such as Fig.~\ref{fig:diag_EMT}b do not contribute to $A$ or $B$; such diagrams contribute
only to the form factors $C$ and $\tilde C$. We have included $\Delta$ isobars in order 
to maintain the correspondence with the large-$N_c$ limit of QCD (see below) \cite{Granados:2016jjl}. 
(II)~By computing the nucleon matrix element of the EMT component $T^{+T}$, Eq.~(\ref{T_Delta}), 
in the LF formulation of ChEFT and constructing the AM density through
Eqs.~(\ref{L_z_from_T}) \cite{Granados:2015rra,Granados:2016jjl}. 
In this first-quantized formulation one follows the evolution of the ChEFT process in LF time $x^+$
(LF time-ordered diagrams). At LO, the only process contributing to $\rho_{A, B}$ and 
$\langle L^z \rangle$ at $b = \mathcal{O}(M_\pi^{-1})$ is Fig.~\ref{fig:diag_EMT}c.
The equivalence of the two methods has been demonstrated at the analytic and numerical level;
the explicit expressions will be given elsewhere \cite{inprep}.
%
%

In the following we present the results in the LF formulation of ChEFT, which is particularly 
suited for illustrating the mechanical properties of the AM densities. In this formulation the 
peripheral nucleon matrix elements are expressed in terms of the LF wave function describing the 
transition of the initial nucleon to the intermediate pion-baryon state ($B = N, \Delta$) 
through the ChEFT interaction (see Fig.~\ref{fig:diag_EMT}c). In momentum representation it is
given by \cite{Granados:2015rra,Granados:2016jjl}
\begin{align}
& \Psi_{N\rightarrow\pi B} (y, \widetilde{\bm{k}}_T, \sigma_B | \sigma)
\equiv 
\frac{\Gamma_{\pi N B} (y, \widetilde{\bm{k}}_T, \sigma_B | \sigma)}
{\Delta\mathcal{M}^2_{N\rightarrow\pi B} (y, \widetilde{\bm{k}}_T)} ,
\label{psi_restframe}
\\[1ex]
& \Delta\mathcal{M}^2_{N\rightarrow\pi B} (y, \widetilde{\bm{k}}_T) \equiv
\frac{\widetilde{\bm{k}}_T^2 + M_\pi^2}{y} + 
\frac{\widetilde{\bm{k}}_T^2 + M_B^2}{\bar y} - M_N^2.
\label{invariant_mass_restframe}
\end{align}
Here $y \equiv k^+/p^+$ is the pion ``plus'' momentum fraction, $\bar y \equiv 1 - y$, and
$\widetilde{\bm{k}}_T$ is the pion transverse momentum in the $\pi B$ configuration.
$\Delta\mathcal{M}^2_{N\rightarrow\pi B}$ is the invariant mass difference between the 
initial $N$ state and the $\pi B$ intermediate state. $\Gamma_{\pi N B}$ is the $\pi N B$ vertex,
which depends on the pion LF momentum in the intermediate state and on the LF helicities of the
intermediate $B$ and initial $N$, $\sigma_B$ and $\sigma$; its specific form is 
given in Ref.~\cite{Granados:2016jjl}. The wave function Eq.~(\ref{invariant_mass_restframe})
is defined in the parametric domain $y = \mathcal{O}(M_\pi / M_N)$ and 
$|\widetilde{\bm{k}}_T| = \mathcal{O}(M_\pi)$, corresponding to soft pions in the nucleon rest frame, 
and is used in this sense only. The peripheral contributions to the transition matrix element of the 
EMT component $T^{+T}$, Eq.~(\ref{T_Delta}), are obtained as
\begin{align}
\langle \bm{T}^{+T} \rangle(\bm{\Delta}_T)
=& \sum_{B = N, \Delta} C_B \, p^+ \int\frac{dy}{y\bar{y}} \int \frac{d^2 k_T}{(2\pi)^3} \;
\sum_{\sigma_B} 
\nonumber \\[1ex]
& \times \Psi_{N\rightarrow\pi B}^* (y, \bm{k}_{T2}, \sigma_B | \sigma) 
\nonumber \\[1ex]
& \times \Psi_{N\rightarrow\pi B}(y, \bm{k}_{T1}; \sigma_B | \sigma) \; \bm{k}_T 
\nonumber
\\[1ex]
& [\bm{k}_{T1, 2} \equiv \bm{k}_T \pm \bar{y} \bm{\Delta}_T/2] .
\end{align}
The expression has the form of an overlap integral between the LF wave functions of the
incoming and outgoing $\pi B$ states ($B = N, \Delta$). $C_B$ is an isospin factor 
($C_N = 3$ and $C_\Delta = 2$; the $\Psi_{N \rightarrow \pi B}$ are normalized such that they 
describe the $p \rightarrow \pi^0 p$ and $p \rightarrow \pi^- \Delta^{++}$ transitions, 
respectively \cite{Granados:2016jjl}).
The first-quantized operator representing the EMT component $T^{+T}$ is given by the
average pion transverse momentum $\bm{k}_T$. Similar results are obtained for the matrix
elements of the other EMT components and the invariant form factors \cite{inprep}.

For expressing the transverse densities it is convenient to use the transverse coordinate
representation of the $N \rightarrow \pi B$ LF wave function,
\begin{align}
\Phi_{N\rightarrow\pi B} (y, \bm{r}_T, \sigma_B | \sigma) \equiv &
\int \frac{d^2\widetilde{k}_{T}}{(2\pi)^2} \, e^{i \bm{r}_T \cdot
\bm{\widetilde{k}}_T} 
\nonumber \\
&\times \, \Psi_{N\rightarrow\pi B} (y, \bm{\widetilde{k}}_T, \sigma_B | \sigma) ,
\label{psi_coordinate}
\end{align}
where $\bm{r}_T$ is the transverse separation of the intermediate $\pi B$ system in the initial
nucleon rest frame \cite{Granados:2016jjl}. In this representation the AM density
Eq.~(\ref{L_z_from_T}) is obtained as 
\begin{align}
\langle L^z \rangle(b)
=& \sum_{B = N, \Delta} \frac{C_B}{4\pi} \int\frac{dy}{y\bar{y}^3} 
\; \sum_{\sigma_B} \; \Phi_{N\rightarrow\pi B}^*(y,\bm{r}_T, \sigma_B | \sigma)
\nonumber \\
&\times \left[\bar{y} \bm{r}_T\times(-i)\frac{\partial}{\partial \bm{r}_T}\right] 
\Phi_{N\rightarrow\pi B} (y,\bm{r}_T, \sigma_B | \sigma)
\nonumber \\[1ex]
& [\bm{r}_T = \bm{b}/\bar{y}, \, \sigma = +1/2].
\label{rholz_wfo}
\end{align}
This result has an appealing form: the first-quantized operator representing the AM is the 
quantum-mechanical orbital AM of the pion in the chiral processes. Equation~(\ref{rholz_wfo}) 
can be used for the physical interpretation and numerical evaluation of the peripheral AM density.
We emphasize that Eq.~(\ref{rholz_wfo}) is an exact representation of the LO ChEFT result and
equivalent to the invariant calculation of the EMT form factors.
\section{Interpretation and evaluation}
%
%
\begin{figure}[t]
\includegraphics[width=0.95\columnwidth]{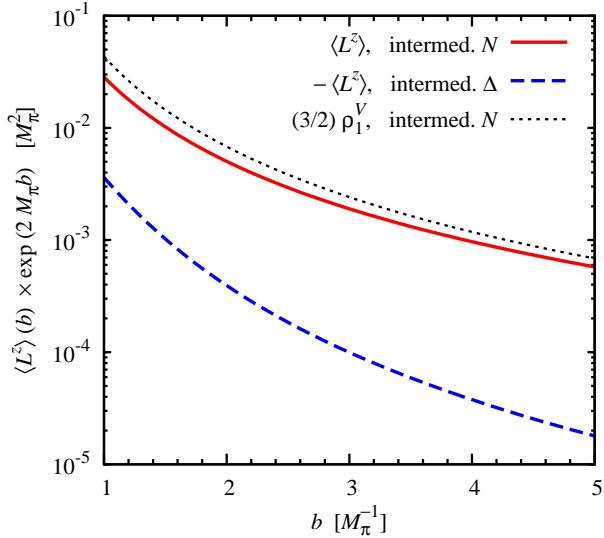}
\caption{Peripheral transverse density of orbital AM in the nucleon obtained from ChEFT, 
Eq.~(\ref{rholz_wfo}). The graph shows the rescaled density 
$\langle L^z \rangle (b) \times \exp (2 M_\pi b)$.
{\it Solid (red) line:} Intermediate nucleon contribution. 
{\it Dashed (blue) line:} Intermediate $\Delta$ contribution with reversed sign 
(the actual values are negative). 
{\it Dotted (black) line:} Isovector charge density $(3/2) \, \rho_1^V(b)$, 
rescaled by the same exponential factor, cf.\ Eq.~(\ref{am_vs_charge_density}).}
\label{fig:OAM_Tden}
\end{figure}
The first-quantized representation Eq.~(\ref{rholz_wfo}) permits a quantum-mechanical interpretation 
of the peripheral AM density (see Fig.~\ref{fig:diag_EMT}d). The initial nucleon with LF helicity 
$\sigma = +\frac{1}{2}$ makes a transition to a $\pi B$ intermediate state ($B = N, \Delta$). 
The peripheral pion in this state has orbital AM; in the state with $B = N$ the allowed values are 
$L^z = \{+1, 0\}$; in the state with $B = \Delta$ they are $L^z = \{+2, +1, 0, -1\}$.
The peripheral AM density in the nucleon is given by the quantum-mechanical AM density of the pion, 
summed over all intermediate states. We emphasize that this interpretation is obtained from the result
of ChEFT, the actual large-distance dynamics of QCD, not from a phenomenological pion cloud model.

The numerical result for the peripheral AM density is shown in Fig.~\ref{fig:OAM_Tden}
(the radial wave functions used in evaluating Eq.~(\ref{rholz_wfo}) are given in
Ref.~\cite{Granados:2016jjl}). The graph shows $\langle L^z\rangle(b)$ as a function of $b$, after
removing the exponential dependence $\sim\exp (-2 M_\pi b)$. One sees that the contribution from
intermediate $N$ states is positive and dominates the peripheral AM density; the contribution from
intermediate $\Delta$ states is negative and numerically small. Note that the signs of the
contributions are in accordance with the quantum-mechanical interpretation.

The ChEFT result for the peripheral AM density has several other interesting properties.
Because the ChEFT calculation includes $N$ and $\Delta$ intermediate states, the peripheral 
AM density has the correct scaling behavior in large-$N_c$ limit of QCD \cite{Granados:2016jjl}; 
see Ref.~\cite{Cohen:1996zz} for a general discussion. In the first-quantized representation 
of Eq.~(\ref{rholz_wfo}) we find that in the large-$N_c$ limit the intermediate $N$ and 
$\Delta$ contributions become equal and opposite
\begin{align}
\langle L^z \rangle [B=\Delta]
\; = \; - \langle L^z \rangle [B = N]
\hspace{2em} [\textrm{large $N_c$}].
\label{large_n}
\end{align}
The cancellation ensures the proper $N_c$ scaling of the total peripheral AM density,
$\langle L^z \rangle(b) = \mathcal{O}(N_c^0)$ at $b \sim M_\pi^{-1} = \mathcal{O}(N_c^0)$.
The quoted $N_c$ scaling refers to the isoscalar AM densities of 
Eqs.~(\ref{spin_decomposition})--(\ref{J_r_from_T}) considered here; the isovector 
quark spin and orbital AM densities exhibit different scaling. We note that the numerical
densities obtained with the physical $N$ and $\Delta$ masses and couplings are far 
from the large-$N_c$ relation Eq.~(\ref{large_n}) (see Fig.~\ref{fig:OAM_Tden}); the same was 
observed in the electromagnetic transverse densities \cite{Granados:2013moa,Granados:2016jjl}.

The first-quantized representation also allows us to compare the transverse AM density to other 
peripheral densities in the nucleon generated by chiral dynamics. Specifically, we can 
relate it to the isovector transverse charge density $\rho_1^V(b)$, which is
associated with the nucleon's isovector Dirac form factor, cf.\ Eq.~(\ref{rho_A_rho_B}). 
Comparison of the first-quantized expressions at the level of the radial wave 
functions \cite{Granados:2016jjl} shows that the densities are approximately related as
\begin{align}
\langle L^z \rangle (b) \; \approx \; \frac{3}{2} \, \rho_1^V(b); 
\label{am_vs_charge_density}
\end{align}
here we neglect the contribution of the $L = 0$ state in $\rho_1^V$, which is $\sim 30\%$ 
at $b = 1 \, M_\pi^{-1}$, as well as the contact term in $\rho_1^V$. 
Note that this approximate relation is specific to LO CHEFT and
should not be used for modeling the densities at distances $\lesssim 1 M_\pi^{-1}$.

Some comments are in order regarding the domain of applicability of the ChEFT results 
for the peripheral densities. Studies of the electromagnetic transverse densities have 
shown that the ChEFT expressions are applicable only at large distances $b \gtrsim 2 M_\pi^{-1}$. 
At smaller distances the densities are strongly modified by $\pi\pi$ rescattering effects in the 
$t$-channel, which enhance the spectral functions of the electromagnetic form factors on the two-pion 
cut compared to the ChEFT result ($\rho$ resonance). A similar enhancement can be expected in the 
spectral functions of the 
EMT form factors $A$ and $B$. The ChEFT results for the AM density, Eq.~(\ref{rholz_wfo}), should 
therefore not be regarded as realistic numerical approximation to the density at distances 
$b \sim 1 M_\pi^{-1}$. However, we expect that the qualitative conclusions based on 
chiral dynamics (dominance of the symmetric part of the EMT, dominance of orbital AM over spin) 
are valid at distances $b \sim 1 M_\pi^{-1}$ and even smaller. 
We note that $\pi\pi$ rescattering effects could be included in the ChEFT calculation of the 
invariant EMT form factors $A$ and $B$ using the unitarity-based methods developed
in Refs.~\cite{Granados:2017cib,Alarcon:2017asr,Alarcon:2017ivh,Alarcon:2017lhg}.
\section{Summary and prospects}
We have studied the transverse densities of partonic AM in the nucleon at peripheral
distances $b = \mathcal{O}(M_\pi^{-1})$, where the structure is governed by chiral dynamics.
The main conclusions are: (a)~At peripheral distances the AM density is mostly orbital,
while the spin density is suppressed. This follows from the general form of the pionic EMT and
axial current operators in chiral dynamics and is a robust conclusion. (b)~In the first-quantized
LF representation of ChEFT at LO, the field-theoretical AM density coincides with the
quantum-mechanical AM density of the soft pions in the chiral processes. This permits a simple
interpretation of the peripheral AM density and relates it to other quantities of peripheral
nucleon structure.

The relevance of the ChEFT results presented here is twofold: (a) They provide insight into
nucleon structure in QCD; (b) they illustrate general properties of the AM densities in the
context of relativistic field theory. Our findings validate the concept of the LF AM densities 
defined in Eqs.~(\ref{L_z_from_T})--(\ref{S_z}) \cite{Adhikari:2016dir,Lorce:2017wkb}. 
The equivalence between the field-theoretical AM density and the quantum-mechanical AM density
is demonstrated specifically for the transverse density at fixed LF time, which has a well-defined
first-quantized representation even for relativistic systems. This suggests that the LF AM density
is a useful concept for nucleon structure studies also at non-exceptional distances.

The studies reported here can be extended in several ways. Using the same methods we can compute
the $T^{++}$ current in the nucleon's periphery, study the associated AM densities 
(Ji's sum rule \cite{Ji:1996nm}), and relate them to the $T^{+T}$ current Eq.~(\ref{Tperpb}) and 
the AM density Eq.~(\ref{L_z_from_T}) studied in the present work \cite{inprep}. We can also compute
the peripheral $T^{+T}$ and $T^{++}$ currents for transverse nucleon polarization and explore 
possible LF densities of transverse AM \cite{Burkardt:2005hp}. The ChEFT methods described here could
also be used to compute nucleon densities at fixed instant time ($x^0 =$ const.) and illustrate their
physical meaning \cite{Polyakov:2002yz}. Finally, the nucleon form factors of the EMT could be 
computed with methods combining ChEFT and dispersion theory, which include $\pi\pi$ rescattering 
effects through unitarity \cite{Granados:2017cib,Alarcon:2017asr,Alarcon:2017ivh,Alarcon:2017lhg}.
\section*{Acknowledgments}
CG acknowledges support by the Gamow Fellowship program of the Columbian College at George
Washington University.
This material is based upon work supported by the U.S.~Department of Energy, 
Office of Science, Office of Nuclear Physics under contract DE-AC05-06OR23177.

\begin{thebibliography}{99}
%
\bibitem{Jaffe:1989jz}
  R.~L.~Jaffe and A.~Manohar,
  Nucl.\ Phys.\ B {\bf 337} (1990) 509,
  doi:10.1016/0550-3213(90)90506-9
%
\bibitem{Ji:1996nm}
  X.~D.~Ji,
  Phys.\ Rev.\ D {\bf 55} (1997) 7114,
  doi:10.1103/PhysRevD.55.7114
%
\bibitem{Leader:2013jra}
  E.~Leader and C.~Lorc\'{e},
  Phys.\ Rept.\  {\bf 541} (2014) no.3,  163,
  doi:10.1016/j.physrep.2014.02.010
%
\bibitem{Lorce:2017wkb}
  C.~Lorc\'{e}, L.~Mantovani and B.~Pasquini,
  Phys.\ Lett.\ B {\bf 776} (2018) 38,
  doi:10.1016/j.physletb.2017.11.018
%
\bibitem{Polyakov:2002yz}
  M.~V.~Polyakov,
  Phys.\ Lett.\ B {\bf 555} (2003) 57,
  doi:10.1016/S0370-2693(03)00036-4
%
\bibitem{Adhikari:2016dir}
  L.~Adhikari and M.~Burkardt,
  Phys.\ Rev.\ D {\bf 94} (2016) no.11,  114021,
  doi:10.1103/PhysRevD.94.114021
%
\bibitem{Gasser:1983yg}
  J.~Gasser and H.~Leutwyler,
  Annals Phys.\  {\bf 158} (1984) 142,
  doi:10.1016/0003-4916(84)90242-2
%
\bibitem{Weinberg:1991um}
  S.~Weinberg,
  Nucl.\ Phys.\ B {\bf 363} (1991) 3,
  doi:10.1016/0550-3213(91)90231-L
%
\bibitem{Bernard:1995dp}
  V.~Bernard, N.~Kaiser and U.~G.~Meissner,
  Int.\ J.\ Mod.\ Phys.\ E {\bf 4} (1995) 193,
  doi:10.1142/S0218301395000092
%
\bibitem{Scherer:2012zzd}
  S.~Scherer and M.~R.~Schindler,
  Lect.\ Notes Phys.\  {\bf 830} (2012) 1
%
\bibitem{Burkardt:2002hr}
  M.~Burkardt,
  Int.\ J.\ Mod.\ Phys.\ A {\bf 18} (2003) 173,
  doi:10.1142/S0217751X03012370
%
\bibitem{Miller:2010nz}
  G.~A.~Miller,
  Ann.\ Rev.\ Nucl.\ Part.\ Sci.\  {\bf 60} (2010) 1,
  doi:10.1146/annurev.nucl.012809.104508
%
\bibitem{Strikman:2010pu}
  M.~Strikman and C.~Weiss,
  Phys.\ Rev.\ C {\bf 82} (2010) 042201,
  doi:10.1103/PhysRevC.82.042201
%
\bibitem{Granados:2013moa}
  C.~Granados and C.~Weiss,
  JHEP {\bf 1401} (2014) 092,
  doi:10.1007/JHEP01(2014)092
%
\bibitem{Granados:2015lxa}
  C.~Granados and C.~Weiss,
  Phys.\ Rev.\ C {\bf 92} (2015) no.2,  025206,
  doi:10.1103/PhysRevC.92.025206
%
\bibitem{Granados:2015rra}
  C.~Granados and C.~Weiss,
  JHEP {\bf 1507} (2015) 170,
  doi:10.1007/JHEP07(2015)170
%
\bibitem{Granados:2016jjl}
  C.~Granados and C.~Weiss,
  JHEP {\bf 1606} (2016) 075,
  doi:10.1007/JHEP06(2016)075
%
\bibitem{inprep}
  C.~Granados and C.~Weiss, in preparation
%
\bibitem{Bakker:2004ib}
  B.~L.~G.~Bakker, E.~Leader and T.~L.~Trueman,
  Phys.\ Rev.\ D {\bf 70} (2004) 114001,
  doi:10.1103/PhysRevD.70.114001
%
\bibitem{Polyakov:2018zvc}
  M.~V.~Polyakov and P.~Schweitzer,
  Int.\ J.\ Mod.\ Phys.\ A {\bf 33} (2018) no.26,  1830025,
  doi:10.1142/S0217751X18300259
%
\bibitem{Brodsky:1997de}
  S.~J.~Brodsky, H.~C.~Pauli and S.~S.~Pinsky,
  Phys.\ Rept.\  {\bf 301} (1998) 299,
  doi:10.1016/S0370-1573(97)00089-6
%
\bibitem{Becher:1999he}
  T.~Becher and H.~Leutwyler,
  Eur.\ Phys.\ J.\ C {\bf 9} (1999) 643,
  doi:10.1007/PL00021673
%
\bibitem{Voloshin:1982eb}
  M.~B.~Voloshin and A.~D.~Dolgov,
  Sov.\ J.\ Nucl.\ Phys.\  {\bf 35} (1982) 120
   [Yad.\ Fiz.\  {\bf 35} (1982) 213]
%
\bibitem{Leutwyler:1989tn}
  H.~Leutwyler and M.~A.~Shifman,
  Phys.\ Lett.\ B {\bf 221} (1989) 384,
  doi:10.1016/0370-2693(89)91730-9
%
\bibitem{Polyakov:1999gs}
  M.~V.~Polyakov and C.~Weiss,
  Phys.\ Rev.\ D {\bf 60} (1999) 114017,
  doi:10.1103/PhysRevD.60.114017
%
\bibitem{Cohen:1996zz}
  T.~D.~Cohen,
  Rev.\ Mod.\ Phys.\  {\bf 68} (1996) 599,
  doi:10.1103/RevModPhys.68.599
%
\bibitem{Granados:2017cib}
  C.~Granados, S.~Leupold and E.~Perotti,
  Eur.\ Phys.\ J.\ A {\bf 53} (2017) no.6,  117,
  doi:10.1140/epja/i2017-12324-4
%
\bibitem{Alarcon:2017asr}
  J.~M.~Alarcón, A.~N.~Hiller Blin, M.~J.~Vicente Vacas and C.~Weiss,
  Nucl.\ Phys.\ A {\bf 964} (2017) 18,
  doi:10.1016/j.nuclphysa.2017.05.002
%
\bibitem{Alarcon:2017ivh}
  J.~M.~Alarcón and C.~Weiss,
  Phys.\ Rev.\ C {\bf 96} (2017) no.5,  055206,
  doi:10.1103/PhysRevC.96.055206
%
\bibitem{Alarcon:2017lhg}
  J.~M.~Alarcón and C.~Weiss,
  Phys.\ Rev.\ C {\bf 97} (2018) no.5,  055203,
  doi:10.1103/PhysRevC.97.055203
%
\bibitem{Burkardt:2005hp}
  M.~Burkardt,
  Phys.\ Rev.\ D {\bf 72} (2005) 094020,
  doi:10.1103/PhysRevD.72.094020
%
\end{thebibliography}
\end{document}